\input{aipcheck}

\documentclass[
    ,final            
  ]
  {aipproc}

\layoutstyle{6x9}

\begin{document}

\title{
The Regge-plus-resonance model for kaon production on
  the proton and the neutron
}


\author{J. Ryckebusch}{
  address={Department of Physics and Astronomy, Ghent University, Belgium}}

\author{L. De Cruz}{
  address={Department of Physics and Astronomy, Ghent University, Belgium},
}

\author{P. Vancraeyveld}{
  address={Department of Physics and Astronomy, Ghent University, Belgium},
}

\author{T. Vrancx}{
  address={Department of Physics and Astronomy, Ghent University, Belgium},
}

\begin{abstract}
  The Regge-plus-resonance (RPR) framework for kaon photoproduction on
  the proton and the neutron is an economical single-channel model
  with very few parameters. Not only does the RPR model allow one to
  extract resonance information from the data, it has predictive
  power. As an example we show that the RPR model makes fair
  predictions for the $p(e,e'K^{+})\Lambda$ and the $n(\gamma,K^{+})\Sigma
  ^{-}$ observables starting from amplitudes optimized for the
  reaction $p(\gamma, K ^{+})\Lambda$ and $p(\gamma,K^{+})\Sigma
  ^{0}$ respectively.
\end{abstract}

\maketitle


\section{Introduction}
These are exciting times for groups working on the modeling of
photoinduced open strangeness production on the nucleon
$N(\gamma,K)Y$. Indeed, over the last couple of years high-quality
data in an extended energy range have become available. The data
include single and double polarization observables. The selfanalyzing
character of the $\Lambda$ facilitates so-called complete measurements
for the $p(\gamma,K^{+})\Lambda$ process. These measurements could
provide a way to determine the four independent amplitudes for kaon
photoproduction at various combinations of the kinematic variables
$(s,t)$.  Modeling of $N(\gamma,K)Y$ reactions poses some real
challenges which stem from the fact that one is dealing with a weak
channel (with cross sections of the order of $ \mu b$) in an energy
range where one expects many overlapping resonances. Accordingly, the
background-resonance separation is even more challenging than in the
pion production channels.

Coupled-channel approaches provide an analysis framework which are
very demanding on human resources and involve a large amount of
parameters. They represent tools to analyze photomeson production data
in the search of $N ^{*}$ and $\Delta ^{*}$ information like masses,
widths and transition form factors, The Regge-plus-resonance (RPR)
framework for $N(\gamma,K)Y$ and $N(e,e'K)Y$ developed by the Ghent
group \cite{RPRlambda,RPRsigma} is an economical single-channel model
with very few parameters. The RPR model is sufficiently simple so that
it can be used as elementary production operator in hypernuclear
calculations for example. In this contribution, we illustrate that
the RPR model has predictive power. First, one can reasonably predict
$p(e,e'K^+)\Lambda$ observables from amplitudes which are optimized
against the $p(\gamma,K^+)\Lambda$ data. Second, fair results for
``neutron'' targets are obtained by appropriately transforming the
amplitudes for kaon photoproduction on the proton.

\section{The RPR model: less can be  more}

\begin{figure}
 \includegraphics[height=.38\textheight]
{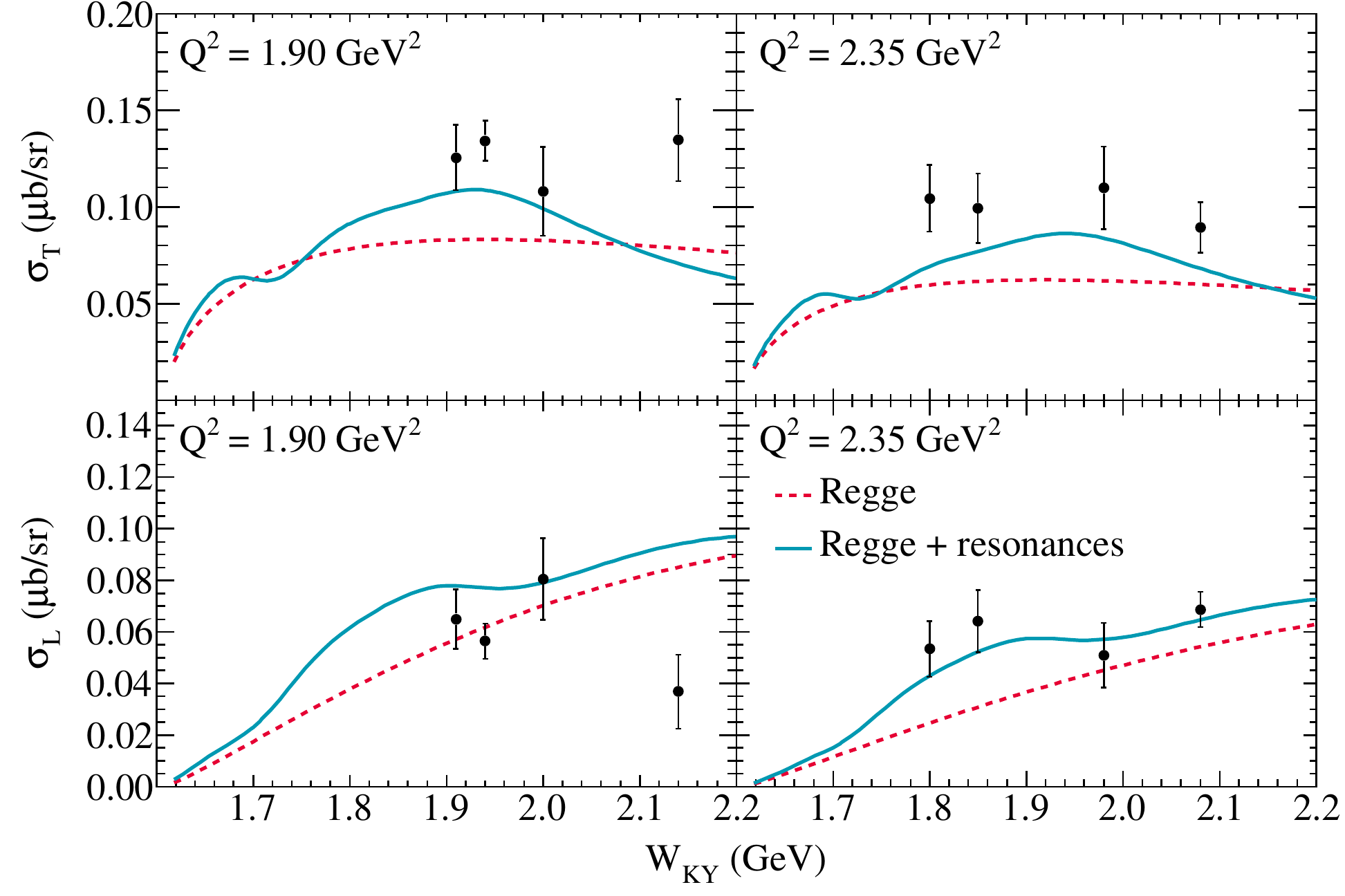}
 \caption{Transverse and longitudinal $p(e,e'K^{+})\Lambda$ cross
   sections as a function of the kaon-hyperon invariant mass $W_{KY}$ for $\cos
   \theta _{K} ^{*} \approx 1 $ and two values of the four-momentum
   transfer $Q^{2}$. The dashed lines show the computed contribution
   from the background diagrams. The solid lines include the
   background and resonant contributions. The data are from
   \cite{Coman}.}
\end{figure}

In the RPR model for $\gamma +p \longrightarrow K ^{+} + \Lambda $ the
background is described by reggeizing the Feynman diagrams involving
pseudoscalar $K^{+}(494)$ and vector $K^{*+}(892)$ meson exchange in
the $t$ channel.  This results in a mere three parameters which fully
determine the background \cite{RPRbayes}.  Adding the electric part of
a reggeized $s$-channel Born diagram ensures that one obtains a
gauge-invariant model \cite{GuidalPhotoProdKandPi}.  In the RPR-2007
version of the model the background parameters are optimized against
the $p(\gamma,K ^{+})\Lambda$ data for $E_{\gamma} \ge 5$~GeV.  These
data were obtained in the sixties and seventies, and it has been
pointed out recently that there may be some normalization issues
\cite{DeyHighE}. The RPR-2011 version of the model uses the
high-energy part of the recent Jefferson Lab data of
Ref.~\cite{McCracken} to optimize the background parameters. The
resonance contributions are added to the reggeized background at the
amplitude level. The resonances constitute the $s$ channel and are
treated in the standard fashion with Feynman diagrams. The RPR-2007
model considers the contribution of the $S_{11}(1650)$,
$P_{11}(1710)$, $P_{13}(1720)$, $P_{13}(1900)$, and $D_{13}(1900)$
resonances to $\gamma +p \longrightarrow K^{+} + \Lambda$. Since 2007
more data have become available and this has allowed us to study also
the role of the $S_{11}(1535)$, $D_{15}(1675)$, $F_{15}(1680)$,
$D_{13}(1700)$, $F_{15}(2000)$, and $P_{11}(1900)$. Not all of these
resonances are identified to play a considerable role. The results of
this analysis of the $p(\gamma,K^+)\Lambda$ data will constitute the
basis of the RPR-2011 model.

The resonance parameters are tuned against the
$p(\gamma,K^{+})\Lambda$ data. For some observables (like the photon
asymmetries for example) one finds that the background describes the
gross feature of the data.  We adopt the view that the $p(e,e '
K^{+})\Lambda$ data provide a stringent test of the predictive power
of the RPR model. A comparison between the RPR-2007 predictions and
recent $p(e,e ' K^{+})\Lambda$ data is shown in Figure~1.  The
measured longitudinal and transverse cross section are of equal
magnitude and exhibit a flat energy dependence.  It is observed that
the three-parameter background provides a fair estimate of the
data. The resonances add some structure and some additional strength
and improve the overall quality of agreement with the data.

\begin{figure}
 \includegraphics[height=.38\textheight]{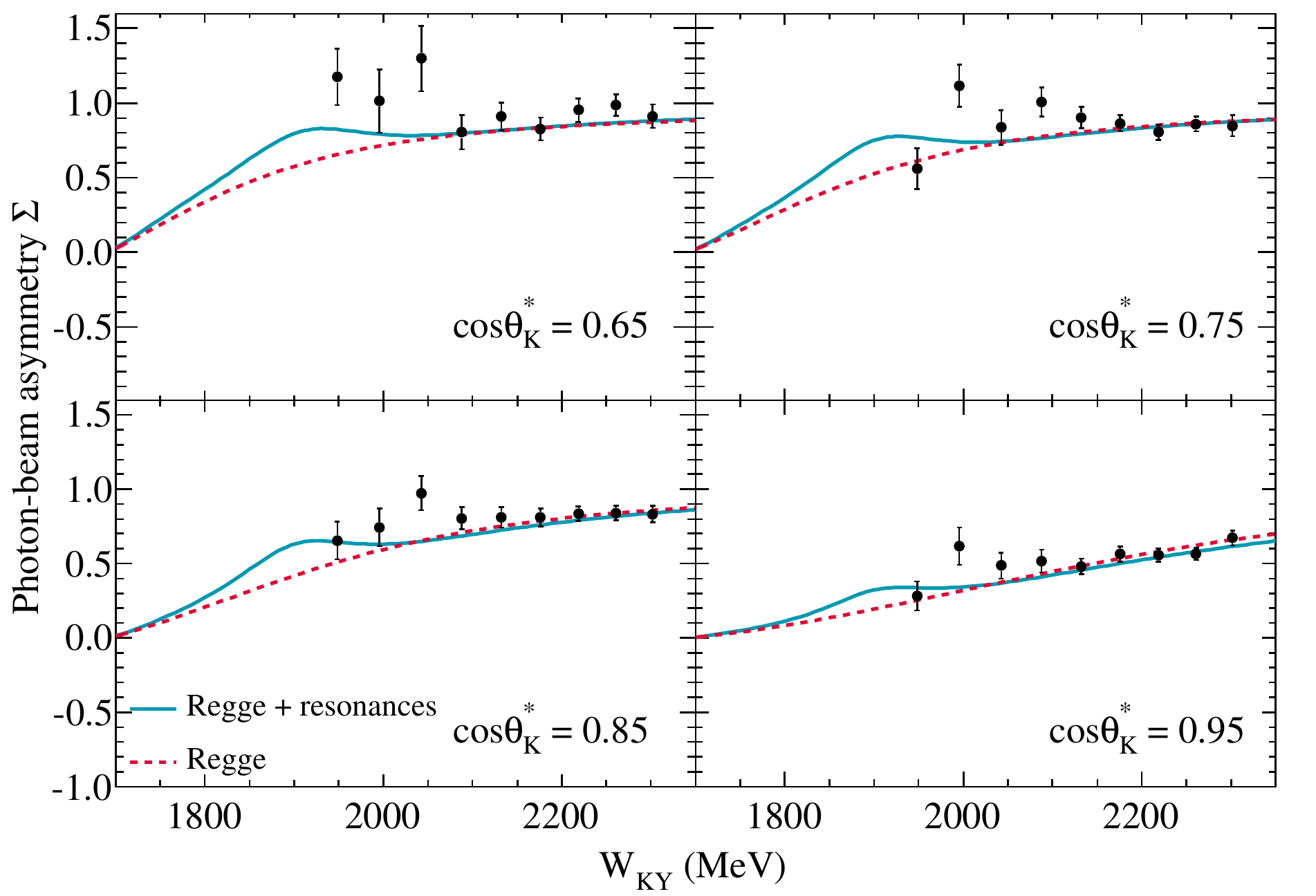}
 \caption{The $n(\gamma,K^+)\Sigma ^{-}$ photon asymmetries as a function
   of the kaon-hyperon invariant mass $W_{KY}$ for various $\cos
   \theta _{K} ^{*}$. The dashed lines show the computed contribution
   from the background diagrams. The solid lines include the
   background and resonant contributions.  The data are
   from \cite{LEPSiso6}. The error bars represent the statistical
   errors.  Systematic errors amount to $\left| \Delta \Sigma \right|
   \approx 0.2$.}
\end{figure}

Strangeness photoproduction involves two $\Lambda$
($p(\gamma,K^+)\Lambda$, $n(\gamma,K^0)\Lambda$) and four $\Sigma$
($p(\gamma,K^+)\Sigma ^{0}$, $p(\gamma,K^0)\Sigma ^{+}$,
$n(\gamma,K^0)\Sigma ^{0}$, $n(\gamma,K^+)\Sigma ^{-}$) channels.  In
order to connect the various channels, it suffices to convert the
coupling constants featuring in the RPR interaction Lagrangians. In
the strong interaction vertices one can fall back on isospin symmetry
to determine the conversion factors for the coupling strengths between
the different channels. In the electromagnetic interaction vertices in
the $s$ channel the coupling strengths $\kappa _{N ^{*}N}$ can be
written in terms of the photocoupling helicity amplitudes
$\mathcal{A}_{J}^{N}$. For example, after equating the proton and
neutron mass, one finds for a $N^{*}$ with $J=\frac{1}{2}$
\begin{displaymath}
\frac {\kappa _{N ^{*}n}} {\kappa _{N ^{*}p}} = 
\frac { \mathcal{A}_{\frac{1}{2}}^{n}} { \mathcal{A}_{\frac{1}{2}}^{p}} \; . 
\end{displaymath}
Experimentally determined values for the helicity amplitudes are
available from analyses like SAID \cite{SAID96} or RPP \cite{pdg2008}
of the pion production data. Unfortunately, in the second and higher
resonance region relevant to kaon photoproduction the extracted values
for $\mathcal{A}_{J}^{N}$ have considerable error bars and are only
available for some selected resonances. In addition, the various
analyses often produce inconsistent results. In Ref.~\cite{RPRneutron}
we have addressed the issue whether the RPR model can predict
$n(\gamma,K^+)\Sigma ^{-}$ observables starting from a limited set of
parameters which are constrained against the $p(\gamma,K^+)\Sigma
^{0}$ reaction channel. In Figure~2 we show the computed RPR result
for the $n(\gamma,K^+)\Sigma ^{-}$ photon asymmetries. The data result
from selecting quasi-free events in $d(\gamma,K^{+}\Sigma ^{-})p$
measurements. The model calculations include both the background and
the resonant contribution and represent predictions which are anchored
to the $p(\gamma,K^+)\Sigma ^{0}$ channel through isospin symmetry. It
is remarkable that with the background diagrams one obtains a fair
description of the angular and energy dependence of the photon
asymmetries. One can conclude that RPR framework allows one to
determine the background contributions to $N(\gamma,K)Y$ in a highly
resilient fashion.  The resonance contributions to the
$n(\gamma,K^+)\Sigma ^{-}$ photon asymmetries are rather moderate.  In
Ref.~\cite{RPRneutron} it is shown that the resonances represent large
corrections to the differential cross sections.  With regard to the
resonance contributions, the mentioned uncertainties on the helicity
amplitudes heavily restrain the predictive power of the RPR model and
other isobar models for $n(\gamma,K)Y$.  The error bars on the
helicity amplitudes, for example, give rise to sizable theoretical
error bands in the computed resonance contributions \cite{RPRneutron}.  

The RPR model provides the elementary production operator for a
recently developed covariant model for $d(\gamma,KY)N$ reactions
\cite{RPRdeuteronBonn}. The role of $YN$ final-state interactions
(FSI) has been evaluated. At low missing momenta, the effect is small
and the deuteron acts as a real neutron target. At high missing
momenta, the $YN$ FSI are large. Under those conditions, the
$d(\gamma,KY)N$ reaction provides a good window on the elusive $YN$
interaction. At high missing momenta, however, the cross sections are
small and the computed $d(\gamma,KY)N$ observables are prone to
uncertainties stemming from the deuteron wave function and the
off-shell extrapolation of the elementary production operator.





\bibliographystyle{aipproc}   


\begin{thebibliography}{12}
\expandafter\ifx\csname natexlab\endcsname\relax\def\natexlab#1{#1}\fi
\providecommand{\enquote}[1]{``#1''}
\expandafter\ifx\csname url\endcsname\relax
  \def\url#1{\texttt{#1}}\fi
\expandafter\ifx\csname urlprefix\endcsname\relax\def\urlprefix{URL }\fi
\providecommand{\eprint}[2][]{\url{#2}}

\bibitem[Corthals et~al.(2006)]{RPRlambda}
T.~Corthals, J.~Ryckebusch, and T.~Van~Cauteren, \emph{Phys.Rev.} \textbf{C73},
  045207 (2006), \eprint{nucl-th/0510056}.

\bibitem[Corthals et~al.(2007)]{RPRsigma}
T.~Corthals, D.~G. Ireland, T.~Van~Cauteren, and J.~Ryckebusch,
  \emph{Phys.Rev.} \textbf{C75}, 045204 (2007), \eprint{nucl-th/0612085}.

\bibitem[Coman et~al.(2010)]{Coman}
M.~Coman, et~al., \emph{Phys.Rev.} \textbf{C81}, 052201(R) (2010),
  \eprint{0911.3943}.

\bibitem[De~Cruz et~al.(2010)]{RPRbayes}
L.~De~Cruz, D.~G. Ireland, P.~Vancraeyveld, and J.~Ryckebusch,
  \emph{Phys.Lett.} \textbf{B694}, 33 (2010), \eprint{1004.0353}.

\bibitem[Guidal et~al.(1997)]{GuidalPhotoProdKandPi}
M.~Guidal, J.~M. Laget, and M.~Vanderhaeghen, \emph{Nucl.Phys.} \textbf{A627},
  645 (1997).

\bibitem[Dey and Meyer(2011)]{DeyHighE}
B.~Dey, and C.~A. Meyer  (2011), \eprint{1106.0479}.

\bibitem[McCracken et~al.(2010)]{McCracken}
M.~McCracken, et~al., \emph{Phys.Rev.} \textbf{C81}, 025201 (2010),
  \eprint{0912.4274}.

\bibitem[Kohri et~al.(2006)]{LEPSiso6}
H.~Kohri, et~al., \emph{Phys.Rev.Lett} \textbf{97}, 082003 (2006),
  \eprint{hep-ex/0602015}.

\bibitem[Arndt et~al.(1996)]{SAID96}
R.~A. Arndt, I.~I. Strakovsky, and R.~L. Workman, \emph{Phys.Rev.}
  \textbf{C53}, 430 (1996), \eprint{nucl-th/9509005}.

\bibitem[Amsler et~al.(2008)]{pdg2008}
C.~Amsler, et~al., \emph{Phys.Lett.} \textbf{B667}, 1 (2008).

\bibitem[Vancraeyveld et~al.(2009)]{RPRneutron}
P.~Vancraeyveld, L.~De~Cruz, J.~Ryckebusch, and T.~Van~Cauteren,
  \emph{Phys.Lett.} \textbf{B681}, 428 (2009), \eprint{0908.0446}.

\bibitem[Vancraeyveld et~al.(2010)]{RPRdeuteronBonn}
P.~Vancraeyveld, L.~De~Cruz, J.~Ryckebusch, and T.~Van~Cauteren, \emph{EPJ Web
  Conf.} \textbf{3}, 03013 (2010), \eprint{0912.2679}.

\end{thebibliography}

\IfFileExists{\jobname.bbl}{}
 {\typeout{}
  \typeout{******************************************}
  \typeout{** Please run "bibtex \jobname" to optain}
  \typeout{** the bibliography and then re-run LaTeX}
  \typeout{** twice to fix the references!}
  \typeout{******************************************}
  \typeout{}
 }

\end{document}